%% file: LLM4Chip.tex
\documentclass[letterpaper]{article} 
\usepackage[]{aaai25}  
\usepackage{times}  
\usepackage{helvet}  
\usepackage{courier}  
\usepackage[hyphens]{url}  
\usepackage{graphicx} 
\usepackage{natbib}  
\usepackage{caption} 
\usepackage{algorithm}
\usepackage{algorithmic}

\usepackage{subfig}
\usepackage[table]{xcolor}         
\usepackage{tcolorbox}
\usepackage{makecell} %

\usepackage{newfloat}
\usepackage{listings}
\DeclareCaptionStyle{ruled}{labelfont=normalfont,labelsep=colon,strut=off} 
\floatstyle{ruled}
\newfloat{listing}{tb}{lst}{}
\floatname{listing}{Listing}
\title{LLM4Laser: Large Language Models Automate the Design of Lasers}
\author{
   Renjie Li\textsuperscript{\rm 1,2}\equalcontrib,
    Ceyao Zhang\textsuperscript{\rm 1}\equalcontrib,
    Sixuan Mao\textsuperscript{\rm 1},
    Xiyuan Zhou\textsuperscript{\rm 1},\\
    Feng Yin\textsuperscript{\rm 1},
    Sergios Theodoridis\textsuperscript{\rm 3},
    Zhaoyu Zhang\textsuperscript{\rm 1}\thanks{Corresponding author: zhangzy@cuhk.edu.cn}
}
\affiliations{
    \textsuperscript{\rm 1}The Chinese University of Hong Kong, Shenzhen\\
    \textsuperscript{\rm 2}University of Illinois Urbana-Champaign, 
     \textsuperscript{\rm 3}University of Athens\\

}

\begin{document}

\maketitle

\begin{abstract}

With the rapid evolution of global autonomous driving technology, the demand for its core sensing hardware, Light Detection and Ranging (LiDAR), is escalating.
As the light source part of the LiDAR system, lasers, particularly the cutting-edge Photonic Crystal Surface Emitting Lasers (PCSEL), have correspondingly attracted extensive research attention.
The conventional manual design and optimization of PCSEL typically require expertise in semiconductor physics and months of dedicated effort to achieve satisfactory results.
While AI-driven approaches can expedite this process, laser designers still need to invest time in learning the AI algorithms involved.
Meanwhile Large Language Models (LLMs), leveraging their powerful reasoning abilities, can effectively comprehend natural language and provide constructive feedback in multi-turn dialogues. They have already demonstrated potential to assist humans in scientific fields such as robotics design and chemical discovery.
A question naturally arises is: Can LLMs transform the lasers design process?
This paper proposes a novel \textbf{human-AI co-design} paradigm to show that LLMs can guide the laser design and optimization process both conceptually and technically. 
Specifically, by simply having conversations, GPT assisted us with writing both Finite Difference Time Domain (FDTD) simulation code and deep reinforcement learning (RL) code to acquire the optimized PCSEL solution, spanning from the proposition of ideas to the realization of algorithms. 
Given that GPT will perform better when given detailed and specific prompts, we break down the PCSEL design problem into a series of sub-problems and converse with GPT by posing open-ended heuristic questions rather than definitive commands. 
We achieved a significant milestone towards \textit{self-driving laboratories}, that is, a fully automated AI-driven pipeline, for laser design and production.

\end{abstract}


\section{Introduction}

Trends of design automation (i.e., human out of the loop) in the integrated circuit (IC), nanotechnology, and semiconductor industries \citep{mirhoseini2021graph,zhang2020neuro,chen2021wafer} are emerging rapidly. Aided by artificial intelligence (AI), machine automation is beginning to replace conventional IC design and fabrication processes involving humans that had existed for over half a century. 
The nanophotonics industry \citep{almeida2004all,altug2006ultrafast,thomson2016roadmap}, however, has not experienced comparable level of automation due to its unique fabrication precision requirements \citep{hocevar2012growth,kim2014all} and relatively complex theoretical models \citep{xie2021higher,zeng2020electrically}. 
As an important example of nanophotonics, the design of Photonic Crystal Surface Emitting Lasers \citep[PCSELs;][] {hirose2014watt,noda2017photonic,yoshida2019double}, 
demands rigorous physical modeling and calculations with Finite Difference Time Domain (FDTD) or Finite Element Analysis (FEA) simulation tools. Moreover, the inverse design and optimization \citep{ma2021deep,molesky2018inverse,so2020deep} of PCSEL typically suffers from three aspects: the one-to-many mapping and non-convexity nature, a high demand for expert knowledge in semiconductor physics and theoretical modeling/simulation (and thus human involvement), and a lack of ready-to-use machine learning algorithms/packages. The above challenges prohibit an end-to-end automated design pipeline for PCSELs and other advanced laser devices alike.

\begin{figure*}[htbp]
  \centering
  \includegraphics[width=0.85\linewidth]{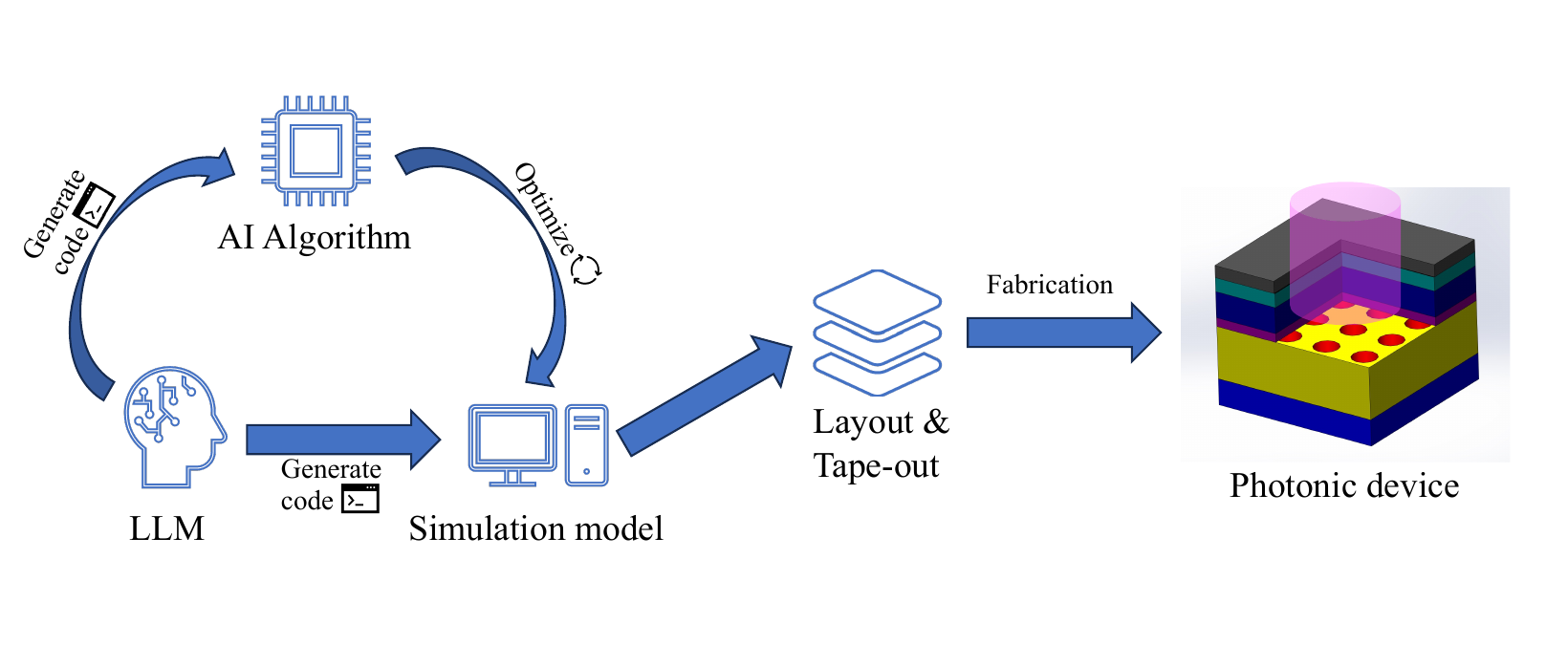}
  \caption{Long-term vision of this work: LLMs for automated PCSEL design and optimization pipeline that enables \textit{self-driving} laboratories. The human facilitator prompts the LLM to generate FDTD code for simulating the PCSEL structure and AI (e.g., reinforcement learning (RL)) code for subsequent optimizations of the PCSEL model. The FDTD code is written with the MIT meep \citep{oskooi2010meep} package. The AI optimization process with RL is built upon an earlier work's L2DO framework \citep{li2023deep}. The final optimized PCSEL design (shown on far right) is then converted to CAD layout and prepared for tape-out and foundry fabrication.}
  \label{fig:self_driving_lab}
\end{figure*}

Luckily, recent advancements in machine learning \citep{goodfellow2016deep,mnih2015human,theodoridis2006pattern} and optimization algorithms \citep{luo2010semidefinite,hale2008fixed,milzarek2014semismooth} have propelled the progress of automated nanophotonics design. Early in the 90s, heuristic, evolutionary \citep{hegde2019photonics}, and gradient-based \citep{zhang2020single} optimization algorithms began to emerge prolifically. Key algorithms include Newton's method \citep{milzarek2014semismooth}, particle swarm \citep{ma2020parameter}, genetic algorithm \citep{ren2021genetic}, Bayesian optimization \citep{shahriari2015taking}, and simulated annealing \citep{bertsimas1993simulated} etc. These algorithms provide a new way of thinking when facing these non-convex optimization problems and lay a solid foundation for continued research. But the challenge remains with heavy human involvement. To solve this, around 2012, researchers proposed deep learning \citep[DL;][]{krizhevsky2012imagenet,goodfellow2016deep} frameworks that take advantage of an abundance of training data and neural network's inference ability. 
These DL models greatly boosted the efficiency of nanophotonic inverse design, pushing the possibility of automated design into a new stage \citep{jiang2021deep,so2020deep,li2022smart,asano2018optimization,li2021deep}.  
Circa 2023, a new DL framework based on RL (e.g., deep Q-learning (DQN) \citep{mnih2015human}), called Learning to Design Optical-Resonators \citep[L2DO;][]{li2023deep}, provides the solution for inverse design of photonic crystal nanocavities without human intervention. 
With two orders of magnitude higher sample efficiency compared to supervised learning,  L2DO has preliminarily realized photonics design automation on an algorithmic level. However, since both the simulation code and DL code in L2DO were still created by the human designer, we were still a distance away from the fully automated photonic design. 

\begin{figure}[tbp]
  \centering
  \includegraphics[width=0.85\linewidth]{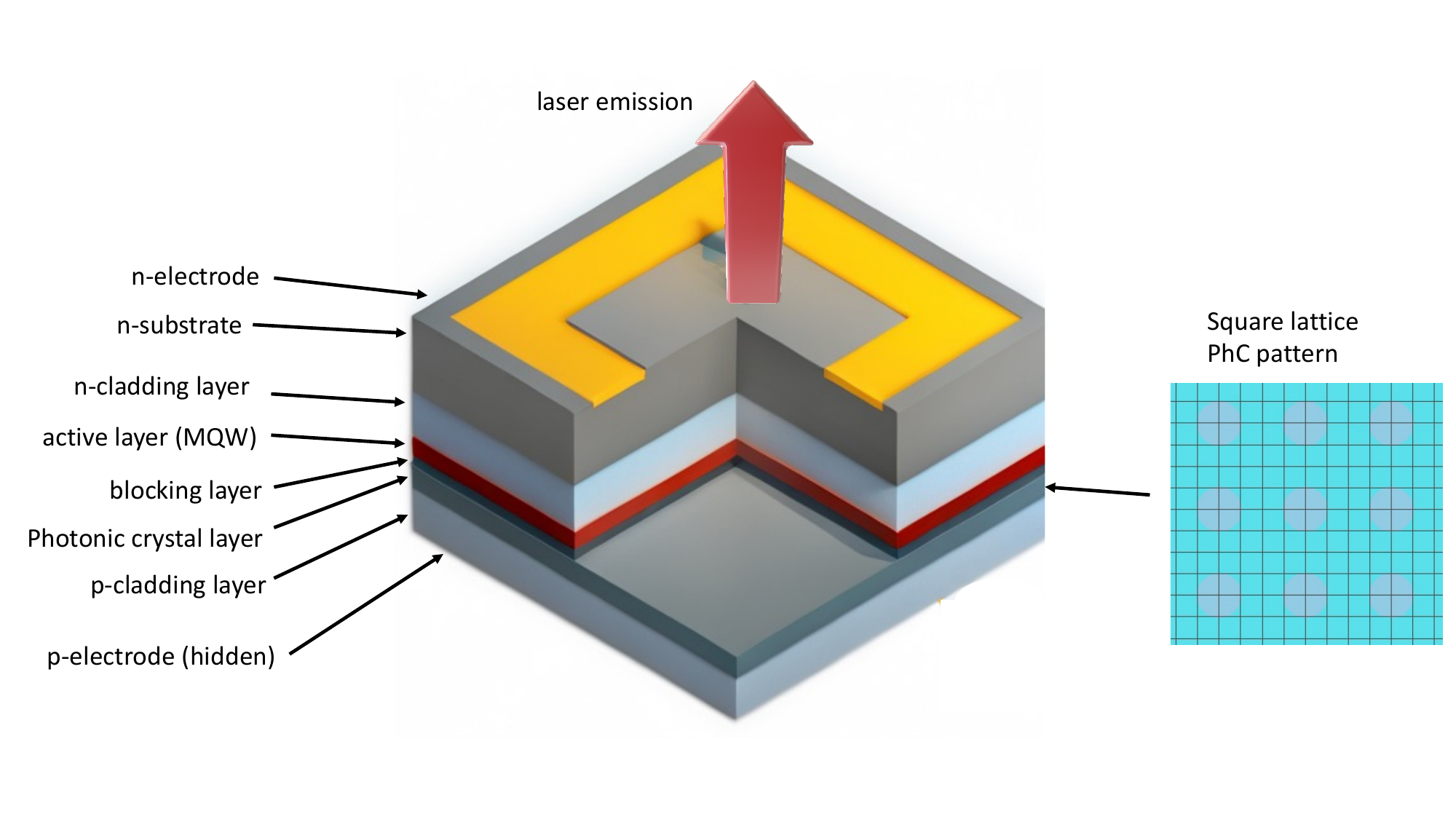}
  \caption{Photonic Crystal Surface Emitting Laser (PCSEL), with abundant applications in sensing, LiDAR, and telecommunications.}
  \label{fig:pcsel}
\end{figure}

Recently, as LLMs has become dominant in the field of AI, a limited handful of researchers have found the potential for using LLMs in hardware design and implementation at an early time. 
In 2020, researchers utilized an improved GPT-2 model called "DAVE" for Verilog code snippets generation and output evaluation \citep{pearce2020dave}, which is a crucial component in the IC design pipeline. A more recent model named "Chip-chat" \citep{blocklove2023chip} came out in 2023, which is an LLM-driven method for IC Verilog code generation and is one of the first wholly-AI-written Hardware Description Language (HDL) for chip tape-out. Meanwhile, LLMs have also contributed significantly to the design and control of robots. Researchers have shown the guidance value of LLM in a robotic gripper design process \citep{stella2023can}, both conceptually and technically. By means of simply conversing with GPT, they successfully designed a robotic gripper capable of reaping the tomato plant. 
Last but not least, LLM-based agent also shows great planning ability in both game AI \citep{wang2023describe, zhang2023proagent} and embodied robotic AI \citep{ahn2022can} tasks.

In this work,  we propose a new \textbf{human-AI co-design} paradigm for PCSELs as illustrated in Figure~\ref{fig:self_driving_lab} and demonstrate the practical implications of LLMs for laser design methodologies. Specifically, we explored and verified the potential of applying LLMs to machine learning-based design and optimization of PCSELs, during which we seek to maintain as little human involvement as possible. By simply having conversations spanning from the proposition of initial ideas to the implementation of final algorithms, GPT-4 assisted us with writing FDTD simulation code and deep RL (e.g. DQN) code to acquire the optimized PCSEL solution. The optimized PCSEL meets the following figure of merit \citep{hirose2014watt}: single-mode, high-beam quality, large-area, and small-divergence angle. A high-level overview of the end-to-end design pipeline is illustrated in Figure~\ref{fig:self_driving_lab}. Given that GPT will perform better when given detailed and specific prompts, we break down the PCSEL design problem into a series of sub-problem modules and converse with GPT by strictly posing open-ended heuristic questions rather than definitive commands. These rules are summarized and proposed as five golden tricks. This paper shows that LLMs, such as ChatGPT, can guide the laser design and optimization processes, on both the conceptual and technical level. All in all, we achieved a significant milestone towards an automated end-to-end laser design and optimization pipeline with minimal human intervention.

\section{Background}

\subsection{Photonic Crystal Surface Emitting Lasers (PCSEL)}
Among lasers, PCSELs \citep{hirose2014watt,noda2017photonic,yoshida2019double} represent an avant-garde technology that integrates the advantages of photonic crystals \citep[PhC;][]{quan2010photonic} and Vertical Cavity Surface Emitting Lasers \citep[VCSELs;][]{chang2000tunable}. PCSELs emit high-quality laser beam vertically and find broad applications in sensing, detection, and telecommunications.

\section{Methods}

\begin{figure*}[t]
  \centering
  \includegraphics[width=0.75\linewidth]{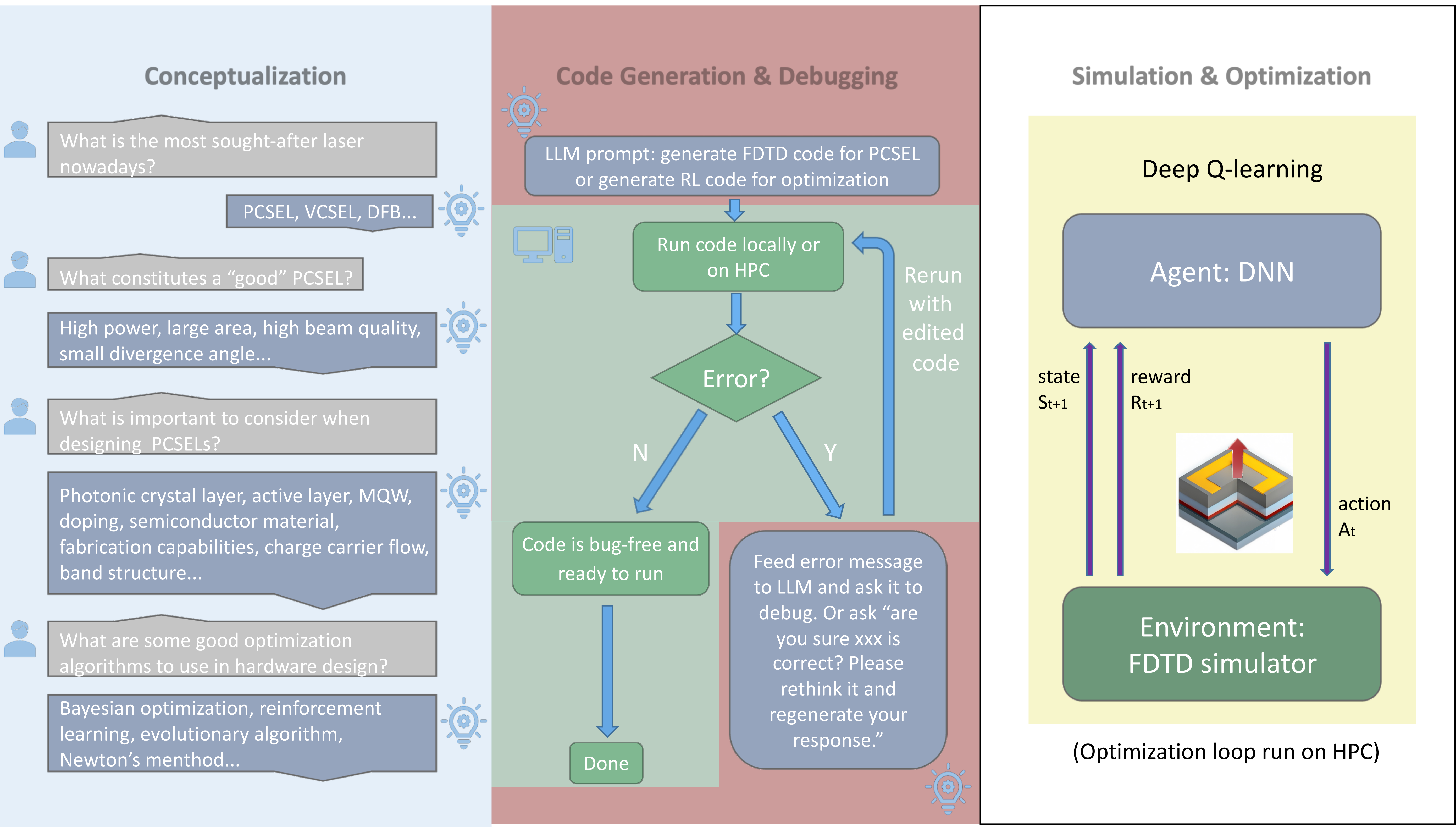}
  \caption{LLM4Laser: A novel Human-AI co-design paradigm for applying LLMs to PCSEL design and optimization. A pictorial overview of the discussions and interactions between the human facilitator and the LLM, with the questions prompted by the human and the answers/solutions provided by the LLM (GPT). The process is divided into three steps: 
  \textbf{left column}: conceptualization, \textbf{middle column}: code generation and debugging, and \textbf{right column}: simulation and optimization. Optimization via DQN is run on high-performance computing (HPC) clusters for improved computational speed and output.}
  \label{fig:llm4pcsel_pipeline}
\end{figure*}

\subsection{Objective overview}

In this article, we investigate the potential of an LLM-based automated PCSEL design and optimization pipeline, as shown in Figure~\ref{fig:self_driving_lab}. 
The target metrics or figures of merit (FOM) of the PCSEL to be satisfied are discussed in Appendix B and C, which were set according to target application specifications and scenarios with full consideration of the physical limits of PCSEL lasing. For example, the wavelength of 1310 nm is important for applications in telecommunications and satellite communications, a high beam quality and a suitably high Q-factor is important for applications like autonomous driving, metal machining/material processing, and medical surgeries, and a small divergence angle is important for achieving high beam quality, small focal spot, and long-distance light propagation. In this initial demonstration, actual tape-out will not be implemented. And due to computational limitations, our PCSEL device has a side length of 2.0 $\mu$m with periodic boundary conditions.

\subsection{Human-AI co-design paradigm}
The \textit{human-AI co-design} paradigm is more clearly illustrated in Figure 3. We divide the design process into three steps: 1) conceptualization, 2) code generation and debugging, and 3) simulation and optimization. Due to the limitation of the status quo of LLMs, the AI agent usually cannot give out the perfect solution all at once. Therefore, the human needs to act as a liaison to help guide/facilitate the design work while simultaneously bearing in mind that excessive human involvement could compromise the integrity of the AI agent's decisions. So for example, if a large portion of the design task is dominated/controlled by humans, it does not reflect the human-AI co-design paradigm and thus should be avoided as much as possible. Overall, in our experience, the design flow should observe the following rules and practice to obtain optimal feedback from the LLM.

First, the whole design process should start with the human providing an open-ended question to GPT, rather than giving definitive commands. This is the beginning of an important conceptualization process, where the human stimulates the LLM to brainstorm and generate creative ideas. For example, you can start a conversation by 
"What is the most sought-after laser nowadays?" or "What are some good optimization algorithms to use in hardware design?" Subsequent conversations will continue by gradually guiding GPT to arrive at a specific solution for this conceptual question. Sample Q\&A rounds are demonstrated in the left column of Figure~\ref{fig:llm4pcsel_pipeline}.
Once we get to the code generation and debugging stage, questions can become more specific and technical, such as "Can you help me design a high-power and large-area PCSEL model using an FDTD algorithm written in Python?" or "Can you help me improve an existing deep-Q learning code implemented with experience replay that's written with PyTorch?"

Second, humans should respect the self-correcting mechanism of GPT rather than directly pointing out the problems/errors it has. In the conversations, there might be times when the answers given by LLMs are self-contradictory or simply wrong. This is attributable to GPT's insufficient understanding of the problem to be solved, which usually happens at the early stages of a conversation. To properly respect the self-correcting mechanism of GPT, one should report the error by responding "you have just mentioned XXX, and I hope you could think twice about this and regenerate your answer" or "are you sure XXX is the correct answer? Please elaborate" rather than "lines XXX and XXX of the code you generated are wrong" or "fix the XXX function/method in the code for me". In the code-generating part, these kinds of mistakes will be especially common. In our experience, it usually takes several conversational iterations before the code finally becomes bug-free and ready to run. Additionally, when debugging the code it is not advised to point out the precise location of errors for GPT; the proper way is to copy the error message from the terminal and let AI do the modification itself. A demonstration of this interactive debugging process is shown in the middle column of Figure~\ref{fig:llm4pcsel_pipeline}. 

Based on the above practical rules, we propose and recommend to the readership the following five golden tricks for successful human-AI co-designs using LLMs:
\begin{enumerate}
    \item \textbf{Open-ended Question Start}: Begin design by posing broad queries to GPT, sparking creative brainstorming and conceptualization. For instance, "What's the latest sought-after laser?" or "Optimization algorithms for hardware design?"
    \item \textbf{Divide and Conquer}: GPT performs better when given detailed questions and information, so break down the design problem into a series of sub-problems.
    \item \textbf{Technical Advancement}: Transition to specific, technical queries during code generation and debugging, like "Develop a high-power PCSEL model using FDTD algorithm in Python."
    \item \textbf{Respect Self-Correction}: Honor GPT's self-correcting capacity by prompting reconsideration of inaccurate answers. Instead of direct corrections, encourage reflection and elaboration for refinement.
    \item \textbf{Effective Debugging Approach}: During code debugging, provide error messages for GPT to address. Refrain from pinpointing error locations; allow GPT to autonomously modify code based on terminal output.
\end{enumerate}

\begin{figure*}[t]
  \centering
  \includegraphics[width=0.75\linewidth]{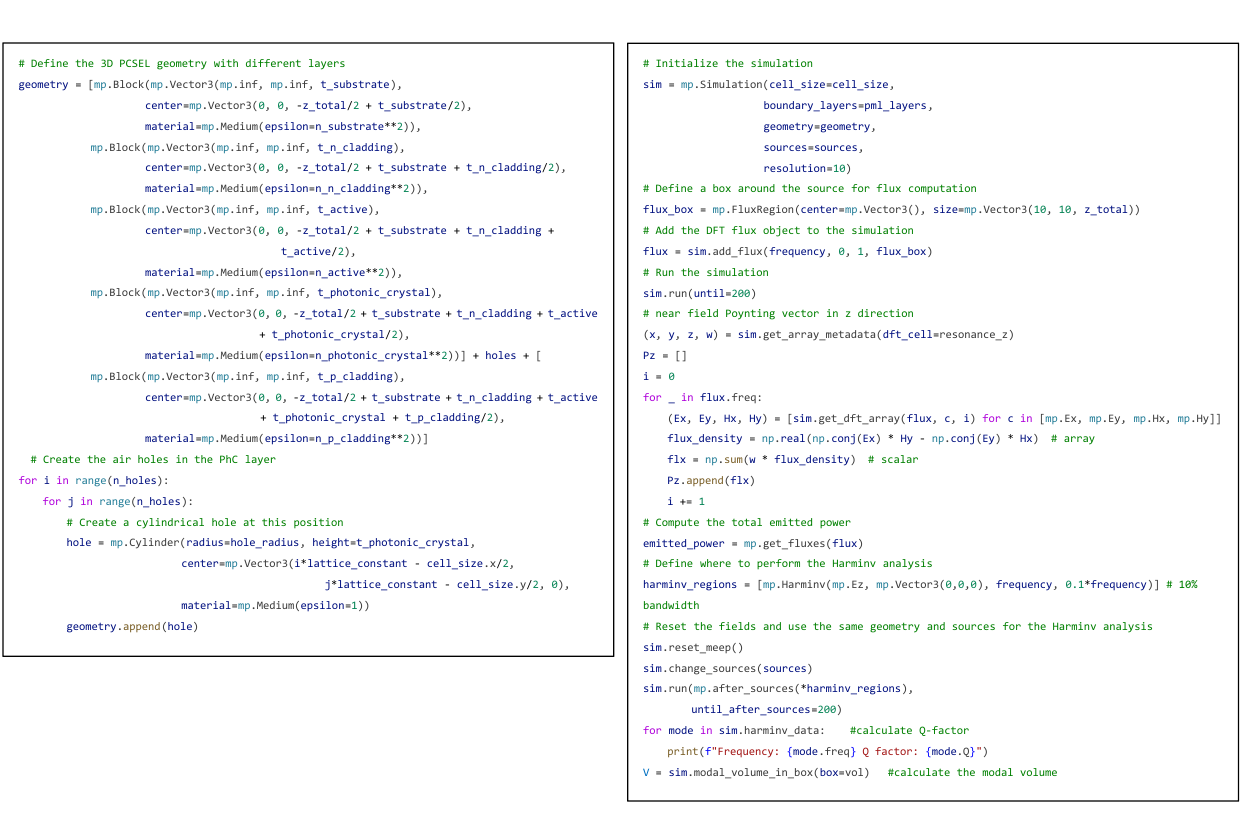}
  \caption{Code generated by GPT-4 for FDTD simulation of PCSEL using the meep package. Left: geometry setup section, right: simulation setup and calculations section. The code shown here is the final version that runs successfully after several rounds of debugging.}
  \label{fig:simulation_code}
\end{figure*}

\subsubsection{Conceptualization with the LLM}
The conceptualization step is for the AI to have a general understanding of the underlying problem so that the AI can choose proper methods and algorithms for more detailed prompts down the road. We kick off the conversation by raising heuristic questions "What constitutes a "good" PCSEL laser?" and "Let us design a PCSEL together, How do you think we should start?", and the LLM provides us with a general design pipeline by saying "here is a general procedure to get started with" where the design problem is broken down into sub-modules such as "understanding the basics of PCSEL", "material selection", "designing photonic crystal structure" etc. Then we take a further step by asking: "I have understood the basics of PCSEL, now what are some important factors to consider when designing PCSELs?" The LLM then points out what to look out for when designing PCSELs. Some sample chats are shown in the left column of Figure~\ref{fig:llm4pcsel_pipeline}. We then prompted another heuristic question about choosing optimization algorithms. GPT-4 eventually provides the answer: "In this case, reinforcement learning and Bayesian optimization might be the most suitable for your problem" after we've clarified our optimization objectives and constraints. When we think that GPT-4 has, for the most part, understood what we are trying to do, we ask it: "Could you generate an appropriate code skeleton according to the above conversation? Please note that the FDTD should be implemented with meep and RL should be with PyTorch." As a result, GPT-4 gives us a code skeleton for FDTD simulation using meep \citep{oskooi2010meep} and DQN using PyTorch, respectively. The problem has now evolved to a matter of expanding these code skeletons to full-blown scripts, meaning that our work is moving to the next stage for code generation.

\subsubsection{Code generation and improvement}
In previous step, we have divided the coding problem into two modules (FDTD and DQN) and obtained the initial code skeletons of both modules. In this step, we will complete, debug, and improve our existing code skeletons, which is an essential step of the whole PCSEL design process. This step is pictorially illustrated in the middle column of Figure~\ref{fig:llm4pcsel_pipeline}.

For generating an FDTD simulation code from a skeleton script, we need to provide the LLM with more specific and concrete physical parameters concerning the PCSEL. For instance, we may require that our PCSEL has five layers, called "n-substrate layer", "n-cladding layer", "active layer", "photonic crystal layer", and "p-cladding layer" respectively, as well as $50\times50$ air holes in the PhC layer. It is worth mentioning that the value of parameters, such as the refractive index and the thickness of each layer, could be randomly initialized due to the DQN optimization process that we are going to implement subsequently. 
In the conversations, we assigned those values in line with the PCSEL model that we have built in the past. In addition, we also provided requirements for boundary conditions, meshing resolution, refractive indices, etc. After we provided the concrete physical parameters, GPT-4 expanded our code by completing the PCSEL geometry and simulation settings, as shown in Figure~\ref{fig:simulation_code}. Additionally, the flux and far-field monitors are added according to our requirement for calculating the emitting power, modal volume, and divergence angle; the Harminv monitor is added for calculating the Q-factor. Some of these monitors and their calculations are shown in Figure~\ref{fig:simulation_code}. 
Now that the FDTD simulation code has been written, the next step is to proceed with debugging and fine-tuning until the code finally runs successfully. As the middle column of Figure~\ref{fig:llm4pcsel_pipeline} and Figure~\ref{fig:debug} shows, we iteratively test-run the code on our local computer and transfer the error messages to GPT for debugging, repeating this process until the code becomes bug-free. In our experience, most bugs can be eliminated within five iterations. 

Next, we generated the DQN code for RL-based optimization of PCSELs (Figure 6). As a core component of the DQN algorithm, we first need an environment to provide the feedback interface.  We adopted OpenAI Gym \citep{brockman2016openai} as the wrapper class for our environment, which is the FDTD simulator that we have generated and fine-tuned. When letting GPT generate the code for the environment, we told GPT what the state space, action space, and reward function are. Further specifications such as the step size of actions, and the upper and lower bounds of state variables are given to GPT as well. Then, with the environment code, we could complete and implement our DQN code. A main DQN script is finished by GPT based on the code skeleton given earlier, considering requirements for the replay buffer, policy DNN, optimizer, loss function, etc. See Figure~\ref{fig:optimization_code}for the core part of the completed DQN script. Note that the DQN script imports the environment class. The next step is to run the DQN code, letting it interact with our FDTD environment and continuously optimize the PCSEL device. 

Just like the FDTD code, the DQN code is then debugged and fine-tuned by iteratively running the code and feeding error messages to GPT. An example of this process is demonstrated in 
Figure~\ref{fig:debug} in Appendix A. 

\subsubsection{Final optimization step with RL}

For the full optimization loop of PCSEL using DQN, refer to Figure~\ref{fig:L2DO}. The main framework is built upon the one proposed in an earlier work \citep{li2023deep}. The objective here is to optimize the existing PCSEL structure such that the target metrics are met. Therefore at each iteration, computed optical attributes (lasing area, Q-factor ...) that are closer to the target metrics will earn a higher reward. The state is defined as the design parameters of PCSEL, whereas the action comprises the changes made to the state at each iteration. The environment, which is a core component of DQN, is realized with the FDTD simulation code that we generated earlier. A detailed explanation of the working theories of DQN, the state and action setup, the agent/policy net, and the reward definition is reserved in the supplementary material. The optimization loop is run on an HPC cluster that has 20 CPU cores and 2 GPUs with CUDA, where each trial takes up to 5 days to run.

\begin{figure*}[htbp]
  \centering
  \includegraphics[width=0.65\linewidth]{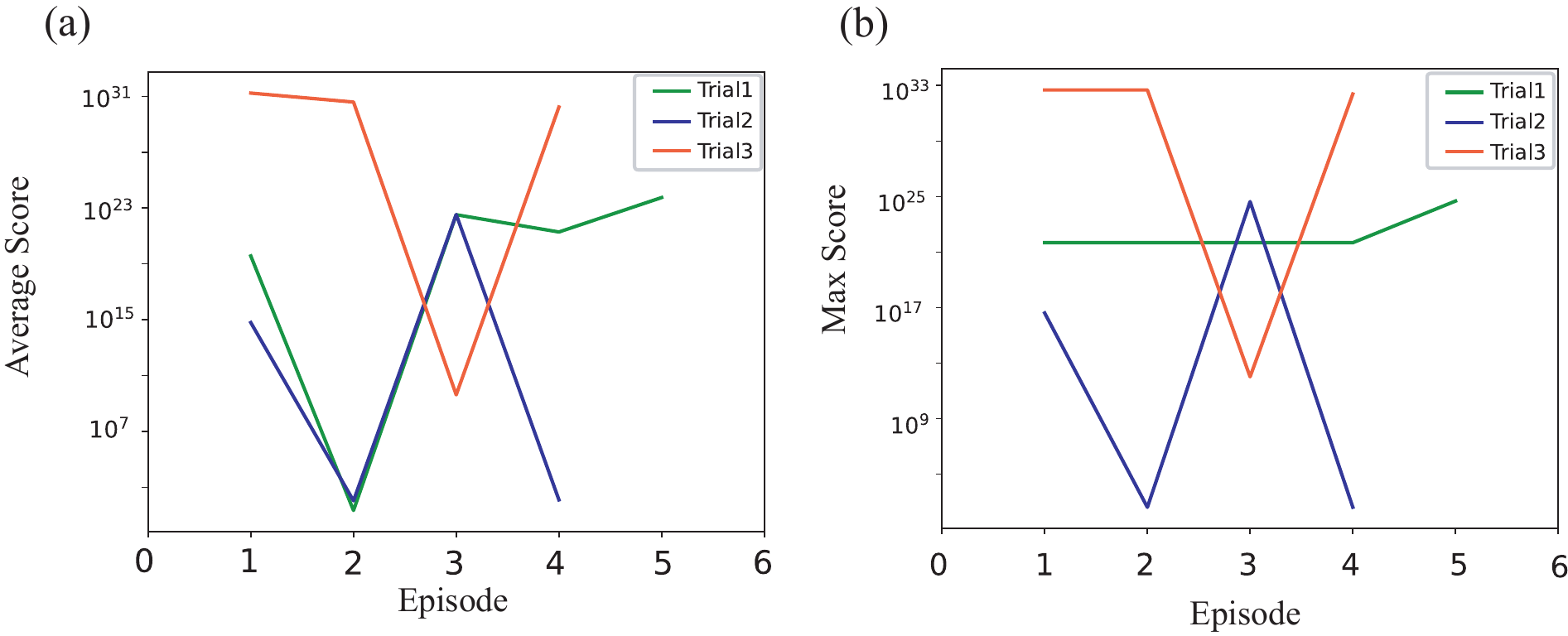}
  \caption{Learning curves of training the DQN to optimize PCSEL, plotted as scores vs. episodes. (a) Average score of each episode; (b) Maximum score of each episode. Each episode contains a horizon of 500 steps. Vertical axes are plotted in log scale.}
  \label{fig:L2DO_training_curve}
\end{figure*}

\begin{table*}[htbp!]
\centering
\begin{tabular}{|c|c|c|}
\hline
\textbf{Metrics}  & \textbf{Optimized values} & \textbf{Literature data \citep{chen2021improvement}}\\ \hline
Operating wavelength (nm) $\uparrow$ & 1383 &  948 \\
Q factor $\uparrow$ & 37000  &  2900  \\
Normalized loss (1/nm) $\downarrow$ & $4.3\times 10^{-7}$ & $7.7\times 10^{-6}$ \\
\hline
\end{tabular}
\caption{Optical attributes of the DQN-optimized PCSEL structure. The rightmost column reports the best literature data (\citep{chen2021improvement}) for a direct metric comparison, which has a operating wavelength of 945 nm. Normalized loss per unit length is additionally reported for a more fair and comprehensive comparison.}
\label{tab:llm4pcsel_results}
\end{table*}

\section{Results and discussion}

Due to the space limitations, we put the details about the optimization of PCSEL via DQN in Appendix B. Figure~\ref{fig:L2DO_training_curve} showcases the learning curves (scores vs. episodes) of training the DQN algorithm to optimize our PCSEL according to the algorithm in Figure~\ref{fig:L2DO}. Three representative trials have been selected to demonstrate the training convergence results. Figure~\ref{fig:L2DO_training_curve}(a) shows the average score of each episode, whereas (b) shows the max score of each episode. Each episode contains a maximum of 500 time steps. Each curve, which represents a complete trial, was trained for 5 days until cut off at the end of the 5th day due to limits on computing resources. 

Using the optimization results illustrated in Figure~\ref{fig:L2DO_training_curve}, the optical attributes of the optimized PCSEL that satisfy the target metrics/figure of merit (FOM) are reported in Table~\ref{tab:llm4pcsel_results}. 
The corresponding set of solved design parameters of the optimized PCSEL is included in Table~\ref{tab:pcsel_optimized_design_params} in suppl. mat.
To better illustrate the advantage of our results, the best PCSEL data from the literature are listed in the rightmost column of Table~\ref{tab:llm4pcsel_results} for a direct comparison of metrics. The literature data were chosen after an exhaustive literature survey \citep{hirose2014watt,noda2017photonic,yoshida2019double,nishimoto2013air,li2023monolithically,inoue2019comprehensive,itoh2020continous,gondaira2016control,kurosaka2008controlling,streifer1977coupled,peng2011coupled,nishimoto2017design,inoue2020design,chen2021improvement} of advancements in PCSEL over the past 10 years, which is fully reported in Table~\ref{tab:full_comparison} of Appendix E.

The device size (side length) of the literature data \citep{chen2021improvement} and our PCSEL is 125.0 $\mu$m and 2.0 $\mu$m, respectively. In this proof-of-concept work, we limited our PCSEL's side length to 2.0 $\mu$m due to insufficient computing resources, as larger models would exponentially increase the simulation time and stall the optimization process. In Table~\ref{tab:llm4pcsel_results}, the calculated wavelength of 1383 nm is within the acceptable tolerance of the target 1310 nm wavelength, being red shifted by 70 nm. The Q factor of our device is over an order of magnitude larger than the literature data, leading to stronger resonance in the PhC layer.
Also, the normalized loss per unit length \citep{kalapala2022scaling} (radiation loss of the resonance mode) of our device is over one order of magnitude smaller than the literature data as seen in Table~\ref{tab:llm4pcsel_results}, which indicates that our device is more energy efficient and lossless in spite of its smaller size. This is also confirmed by the larger Q-factor of our device. However, since we used an infinitely large simulation model in FDTD (Bloch boundary conditions), divergence angle and lasing area are not comparable to the literature data at the moment and we will include these metrics in future endeavors. All in all, we can conclude that the optimized attributes shown in Table~\ref{tab:llm4pcsel_results} have satisfied and even exceeded the target metrics set by us. 

Here, we propose several design techniques to further lower the divergence angle of PCSELs: vary the shape of the lattice (hexagonal vs. square), vary the shape of air holes (triangular vs. circular), increase the device size, and use double-lattice PhC structure \citep{yoshida2019double}. Analytically, according to \citet{wang2022symmetry}, the divergence angle can be expressed as  :  
$$ \theta = \frac{m \lambda}{L}, $$
where $m$ is the coefficient which varies with different structures of the PhC lattice such as the shape of lattice and air holes and the period number of lattice etc., $\lambda$ is the resonant wavelength, and $L$ is the size of PhC slab. Additionally, the divergence angle of PCSEL can be described by the eigenstates in momentum which dictates that the in-plane wave-vector should be close to $q\pi/L$ \citep{chen2022analytical}. Here, $q$ is a coefficient similar to $m$. Therefore, the divergence angle of PCSEL is determined by multiple complex parameters and we will subsequently demonstrate the effect of these variations on the PCSEL performance. In addition, since RL is known to be sample-inefficient and hard to train, we will explore other algorithms such as Bayesian optimization (BO) \citep{shahriari2015taking} that are considerably more efficient and lightweight. BO, as a black-box optimization algorithm suitable for expensive environments, could potentially speed up the optimization process and produce better results. 

Last not but least, we experimented with another latest LLM called Llama2 \citep{touvron2023llama}, which was released by Meta AI in July 2023. As a lightweight (70 billion parameters) and open-source LLM, Llama2 emerges as an attractive alternative to ChatGPT. Using the same conversations, we generated the Meep FDTD code and the DQN code in Llama2, which are shown and analyzed in Appendix D. We conclude that, overall, GPT delivered superior performance and dominated Llama2 in terms of question understanding, idea brainstorming, code generation, error self-correcting, etc. This performance gap, of course, can be attributed to the enormous number of parameters (1.8 trillion) and number of training hours that GPT-4 has in its possession. Nonetheless, for those who wish to work with open-source and free-of-charge LLMs, Llama2 is still a decent choice to start with.

\section{Conclusion}

In this paper, we introduced a novel human-AI co-design paradigm for PCSELs, showcasing the broader relevance of LLMs in laser design scenarios. We systematically explored the application of LLMs in machine learning-driven design and optimization of PCSELs, aiming for minimal human intervention. Through ordinary conversational interactions, ranging from initial concept proposals to final algorithm implementation, GPT-4 aided in crafting FDTD simulation and deep reinforcement learning (e.g. DQN) code to achieve an optimized PCSEL solution meeting criteria like single-mode operation, high beam quality, large area, and narrow divergence angle. This paradigm successfully addressed three major challenges faced by state-of-the-art deep learning-enabled inverse design methods: 1) the fundamental one-to-many mapping or the non-convex issue; 2) heavy human involvement for technical input; and 3) shortage of ready-to-use machine learning methods. The design process involves breaking down the design problem into modular sub-problems and heuristically prompting GPT to answer open-ended questions, among several other golden tricks we summarized and recommended to the audience. We hope that these golden tricks can serve as a general guideline for anyone who wishes to benefit from LLMs' power in hardware design. Our results demonstrate that LLMs, such as ChatGPT, can effectively guide laser design and optimization, both conceptually and technically. In the end, we proposed several future research directions and showcased a comparison to meta's Llama2. We will subsequently pursue these future thrusts to deliver better optimization results. Overall, we mark a significant step toward an AI-empowered automated, end-to-end nanophotonic design and optimization pipeline.




\bibliography{aaai25.bib}

\input{supp}

\end{document}

%% file: supp.tex
\clearpage
\onecolumn

\section*{LLM4Laser: Large Language Models Automate the Design of Lasers}
\section*{Supplementary Material}

\subsection{Appendix A: The debugging demonstrations by GPT}\label{appd:debug}
The code generated by GPT for RL is shown in Figure 6 and the debugging iterations with GPT is shown in Figure 7. 
\begin{figure*}[h]
  \centering
  \includegraphics[width=0.75\linewidth]{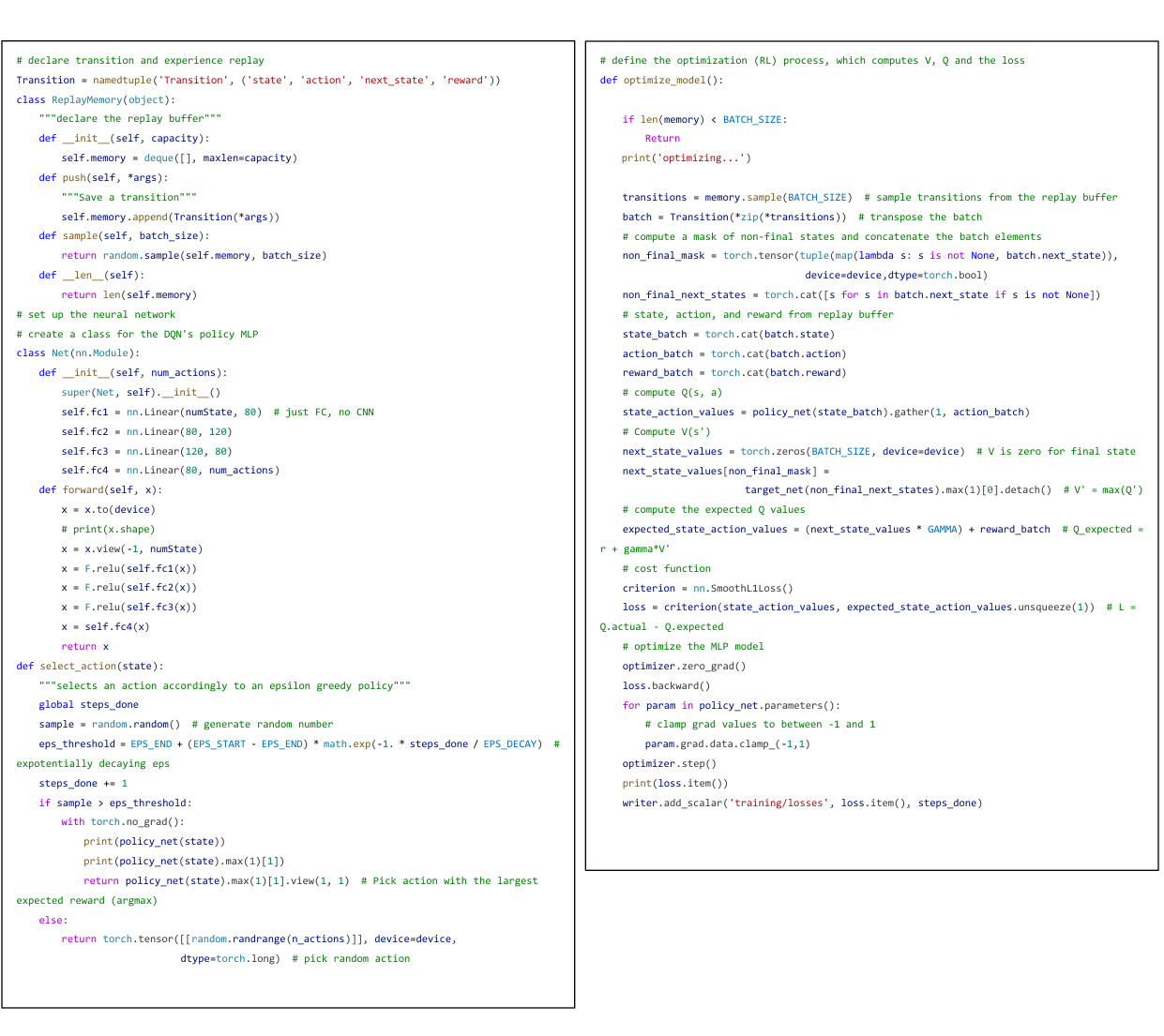}
  \caption{RL code generated by GPT-4 for optimization of PCSEL using DQN. Left: replay buffer, policy network, and action selection setup, right: core part of DQN where the value functions, action-value functions, and loss are computed. The code shown here is the final version that runs successfully after several rounds of debugging with human.}
  \label{fig:optimization_code}
\end{figure*}

\begin{figure*}[htbp]
  \centering
  \subfloat[]{\includegraphics[width=0.45\textwidth]{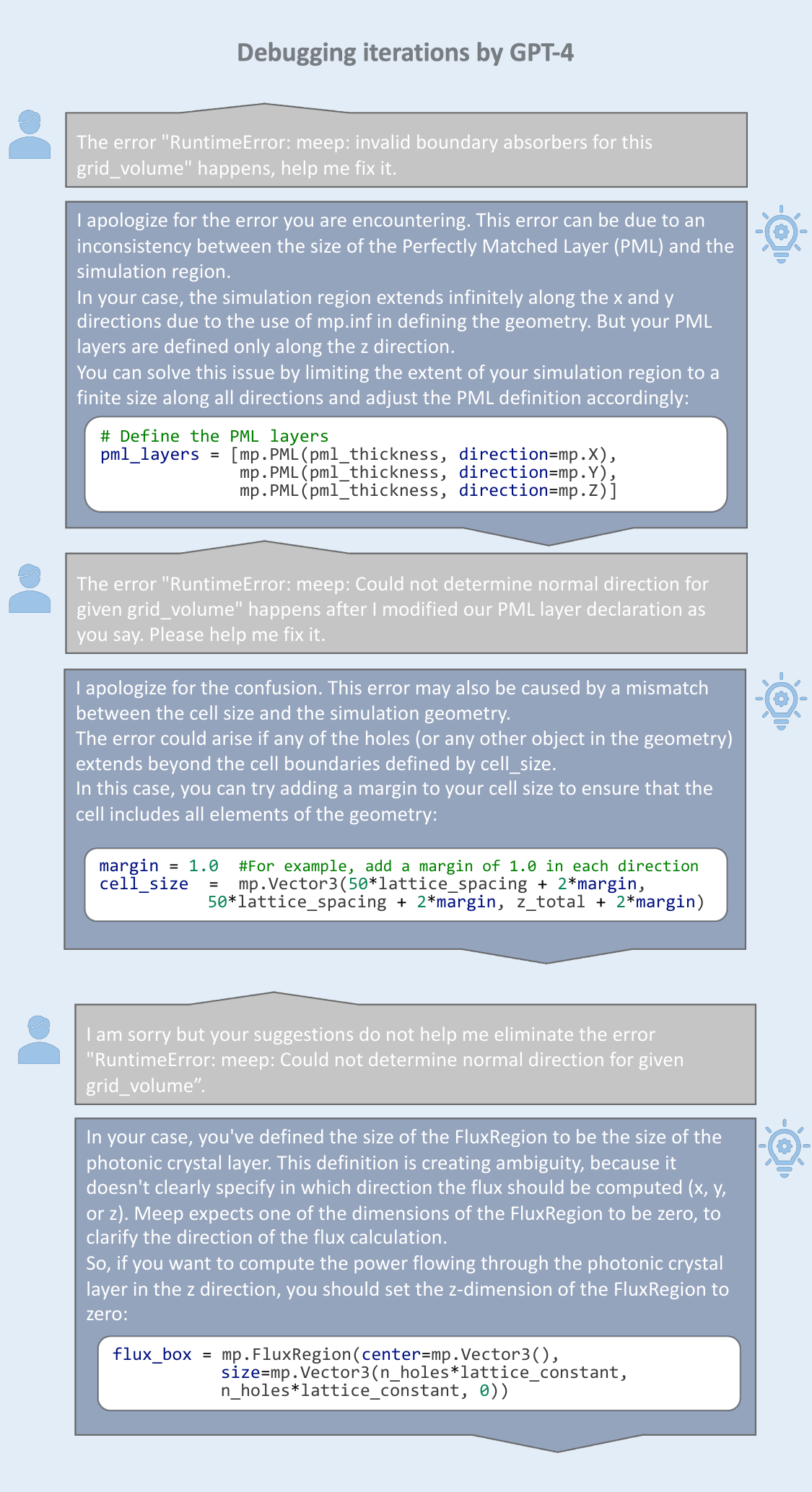}\label{fig:debug_meep}}
  \hfill
  \subfloat[]{\includegraphics[width=0.45\textwidth]{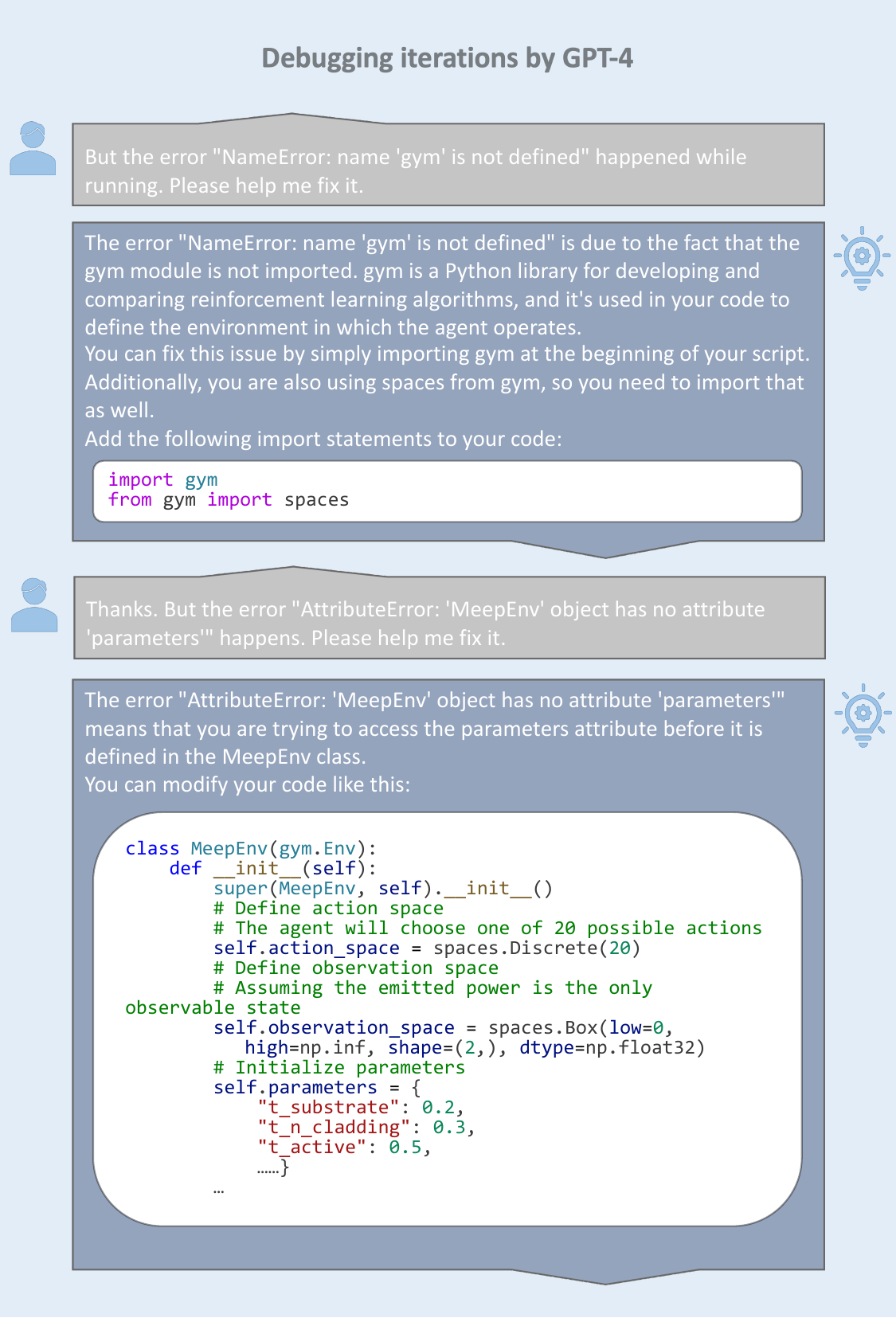}\label{fig:debug_RL}}
  \caption{a) Debugging iterations of the FDTD code by conversing with GPT-4. All bugs/errors in the code demonstrated in Figure~\ref{fig:simulation_code} were cleared out within 5 iterations; 
  b) Debugging iterations of the DQN code by conversing with GPT-4. All bugs/errors in the code demonstrated in Figure~\ref{fig:optimization_code} were cleared out within 5 iterations.}
  \label{fig:debug}
\end{figure*}



\subsection{Appendix B: The details about the optimization of PCSEL via DQN}

\begin{figure*}
  \centering
  \includegraphics[width=0.80\linewidth]{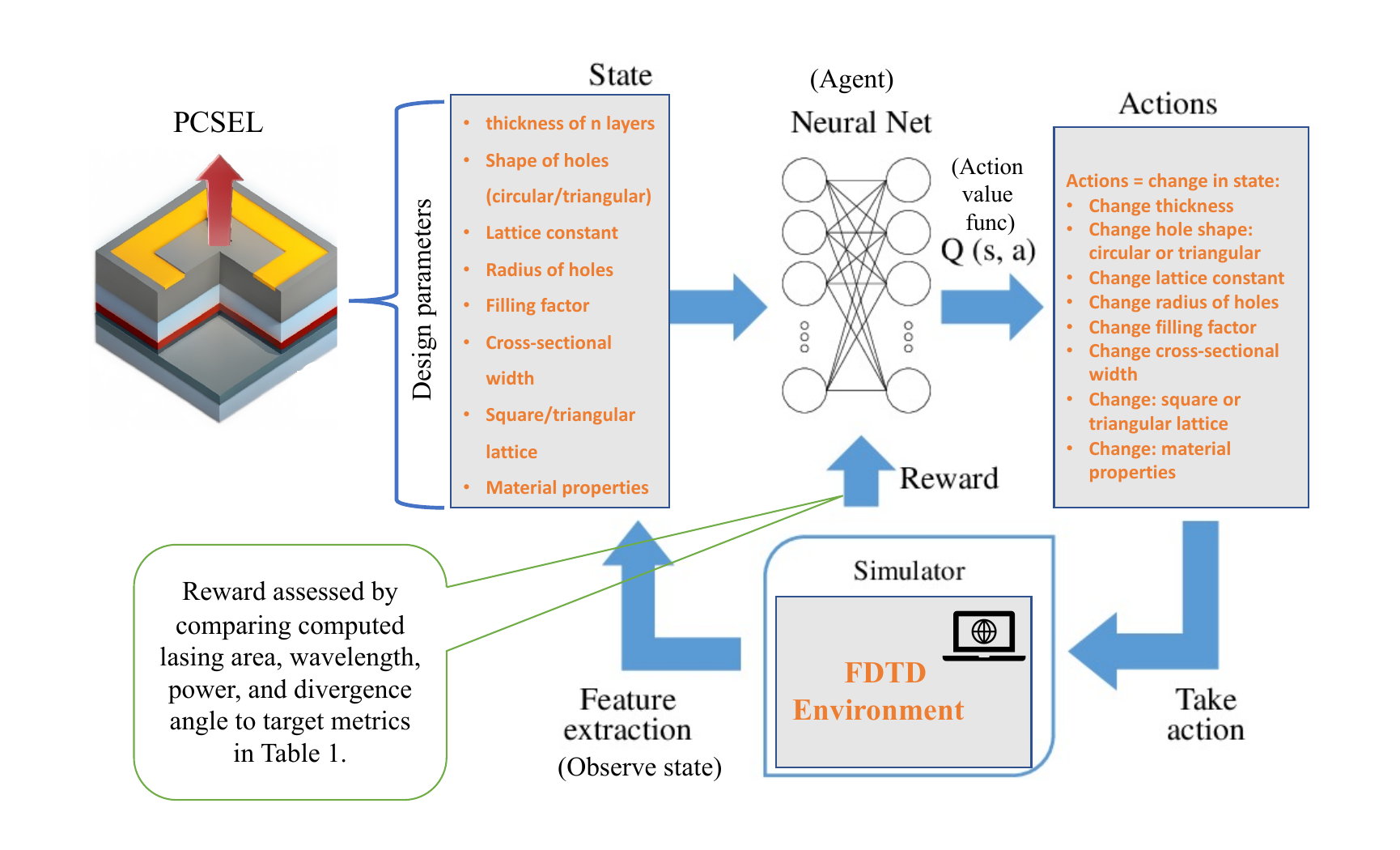}
  \caption{Optimization of PCSEL via reinforcement learning (DQN), where the objective is to meet those target metric/FOM listed in Table~\ref{tab:pcsel_target_metric}. This is a detailed expanded version of the right column of Figure~\ref{fig:llm4pcsel_pipeline}, including the state and action parameters, the reward definitions, the FDTD environment, and the agent DNN. Q(s, a) is action-value functions predicted by the agent DNN.}
  \label{fig:L2DO}
\end{figure*}

\textbf{Deep Q-learning.}
Deep Q-Learning (DQN) is a type of reinforcement learning algorithm that uses a deep neural network to approximate the action-value (Q) function, which is used to determine the optimal action to take in a given state. The Q-function represents the expected cumulative reward of taking a certain action in a certain state and following a certain policy. In Q-Learning, the Q-function is updated iteratively as the agent interacts with the environment. It is the same in Deep Q-Learning. But instead of changing the Q value in a map directly, Deep Q-learning changes the weight in the neural network. Actually  the learning process needs 2 neural networks, called the policy network and target network, respectively. These networks have the same architecture but different parameter weights. Every N steps, the weights from the main network are copied to the target network. In the action selection process, we apply a strategy called Epsilon Greedy Exploration, which can be described as follows: 
1) At every time step when it’s time to choose an action, roll a dice. 
2) If the dice has a probability less than epsilon, choose a random action. 
3) Otherwise take the best known action at the agent’s current state. At the very beginning the epsilon is very large, which means the agent needs to explore the environment and get some weights updated (all the weights are initialized as 0). As the epsilon is decreasing, the agent starts to try existing known good actions more and more. Using both of these networks leads to more stability in the learning process and helps the algorithm to learn more effectively. And about the weight update of the DQN, we use Bellman's Equation, which is shown below:
\begin{equation}
        V_\pi(s) = E_\pi [R_{t+1}+\gamma V_\pi(s_{t+1})|s_t=s)
\end{equation}

A concise introduction to the DQN algorithm is presented in the following text. 
DQN introduces a groundbreaking concept by integrating a policy (action-value) network $Q(s, a)$ with a target network $Q'(s, a)$, leading to substantial improvements in training performance. Initially, $Q'(s, a)$ is set to be an exact replica of $Q(s, a)$, sharing the same parameters. 
parameter C denotes the number of steps before updating the target action-value function $Q'(s, a)$ with the values from $Q(s, a)$, often referred to as the "freeze time." This freezing mechanism has proven to be highly beneficial in enhancing convergence stability and reducing policy oscillations.


Another crucial aspect of DQN is the implementation of experience replay \cite{mnih2015human}, facilitated through a replay buffer denoted as D. At each time step, the agent's transition $(s, a, r, s')$, also known as experiences, is stored in a pre-allocated array called D. During the policy update using SGD, experiences from D are randomly selected as training samples. Experience replay offers several advantages: firstly, it allows past experiences to be reused in numerous future gradient updates, leading to enhanced sample efficiency and potentially faster convergence; secondly, as consecutive samples are often correlated and exhibit similar distributions, this can cause the learning process to get stuck at local minima. By randomizing these samples, the data correlation is broken, enabling a more diverse data distribution. Experience replay can also smooth out learning curves and alleviate oscillations or even divergence during training.

Next, $\epsilon$-greedy plays a crucial role in DQN. Given that DQN operates in an off-policy manner, it directly estimates actions using the greedy policy $a = \arg\max_{a'}Q(s, a')$. However, to strike a balance between exploration and exploitation and enable the agent to explore a broader range of the state space, $\epsilon$-greedy dynamically adjusts this policy. The agent follows the $\epsilon$-greedy policy based on $Q(s, a)$, as outlined in Algorithm 1. Practically, the $\epsilon$-greedy policy mostly adheres to the greedy policy, selecting the action with the highest estimated Q-value with a probability of $1-\epsilon$. However, with a probability of $\epsilon$, the agent selects a random action to promote exploration. In this work, the initial and final values of $\epsilon$ are set to 0.90 and 0.05, respectively.

The defined objective function, also known as the loss function, for this problem, is as follows: 
\begin{equation}
    L(\theta) = \mathrm{E} [(r + \gamma \max_{a'} Q'(s', a) - Q(s, a))^2]
\end{equation}
To optimize the loss function $L(\theta)$ presented in Equation 1, stochastic gradient descent is employed. In our conducted experiments utilizing DQN, we utilized the RMSprop optimizer with minibatches of size 32 and a learning rate of 0.00025. In Equation 1, the variable $r$ represents the reward, and we set the discount factor $\gamma$ to 0.99. This choice of $\gamma$ allows us to estimate the cumulative return defined at a future time point \textit{T}. The cumulative return is computed as the discounted sum of all future rewards:
\begin{equation}
   R = \sum_{t}^{T}\gamma^t r_t
\end{equation}
Through the optimization of the loss function $L(\theta)$ as defined in Equation 1, the primary goal is to maximize the cumulative return $R$ as expressed in Equation 2. By achieving this objective, we aim to identify and obtain the optimal action that we are seeking in our context.

\textbf{The state and action setup}
In the paper we have briefly introduced our PCSEL structure and some elements, here we present a detailed information about it: 

Among all these design parameters, we choose 10 to be optimized, which are the hole radius, the lattice spacing, the thickness of all five layers, the reflective index of n-cladding layer (we let the refractive index of p-cladding layer change with n-cladding layer), the active layer and the substrate layer. Each parameter to be optimized has two changing directions. Thus we have 20 discrete action space over all. For change of thickness, we let the step to be 0.0005 or minus 0.0005. For the change of lattice spacing and reflective index, we set the step of 0.0005 or minus 0.0005. For hole radius, we let the change to be 0.0001 each time. The following table shows the boundary of the parameters to be changed.\par

Table~\ref{tab:pcsel_desgin_params} presents the state spaces and action spaces in our DQN algorithm. For this proof-of-concept work, the PCSEL is designed to have 10 states and 20 discrete actions, striking a balance between a manageable parameter space and limiting the overall training time. It is worth noting that more comprehensive investigations can be chosen for higher-order state-action spaces in future studies. The state space, representing a subset of design parameters, encompasses geometric parameters of the PCSEL such as thicknesses and lattice spacing. These states serve as inputs to the policy network. On the other hand, the action space is constructed by incrementing or decrementing each state by a fixed step size of 0.0001 or 0.0005, as outlined in Table~\ref{tab:pcsel_desgin_params}. To determine the optimal action, the policy network predicts an action, which is then utilized to update the state in the FDTD environment. Subsequently, the environment calculates the associated rewards based on the updated state. For instance, if the current state is $s = 0.3$ and the action is $a = +0.0001$, the environment will yield the next state as $s' = 0.3 + 0.0001 = 0.3001$, along with the associated reward as $rew'$. The output reward and the resulting next state are fed back into the policy to initiate the subsequent iteration, and concurrently, they are used to update the policy network. 
Past states and rewards are typically stored in a replay buffer to be utilized later and to mitigate sample correlations \cite{mnih2015human}. If, after multiple action steps, the value of the current state exceeds the boundaries defined by the Min and Max values specified in Table~\ref{tab:pcsel_desgin_params}, the current episode will be forcibly terminated, and a new episode will begin. For a more comprehensive and detailed understanding of the DQN's mechanism, additional elaborations and in-depth information can be found in the original DQN paper \cite{mnih2015human}, enabling readers to gain a deeper insight into the process.

\begin{table}[htbp]

\centering
\begin{tabular}{|c|c|c|}
\hline
\multicolumn{3}{|c|}{\textbf{PCSEL design parameters}}\\\hline
\textbf{State space} & \textbf{Min}  &  \textbf{Max}  \\ \hline
substrate layer thickness & -0.3 um&  0.3 um \\
n cladding layer thickness & -0.3 um&  0.3 um \\
active layer thickness & -0.3 um &  0.3 um\\
PhC layer thickness & -0.3 um &  0.3 um \\
p cladding layer thickness & -0.3 um&  0.3 um  \\
refractive index of substrate layer & -0.15 &  0.15  \\
refractive index of n cladding /p cladding layer & -0.15 &  0.15  \\
refractive index of active layer & -0.15 & 0.15  \\
lattice spacing &  -0.1 um&  0.1  um\\
hole radius & -0.1 um &  0.1 um \\ \hline \hline
\textbf{Action space}& \textbf{Total No. of actions}& \textbf{Action type}\\ \hline
each state $\pm$ a step size & 20  &  Discrete \\
\hline 
\end{tabular}
\caption{State space and action space of this DQN-based optimization of PCSEL. State variables are net changes in design parameters (i.e. state = $\Delta$design parameter)}
\label{tab:pcsel_desgin_params}
\end{table}

\textbf{Reward formulation}

To assess the quality of the PCSEL, we set four criteria, which are resonant wavelength, emitting power (or Q-factor), lasing area, and divergence angle. Each criteria is going to be converted to a reward, and the score is the weighted normalized sum of all four rewards.   

Equation 4-8 defines the reward and how it's related to the target optical responses as laid out in the main text:
\begin{equation}
    rew_{1} = 1 - |\lambda^\ast - \lambda| /\lambda^\ast
\end{equation}

\begin{equation}
    rew_{2} = 1-(area^\ast-area) / area^\ast
\end{equation}

\begin{equation}
    rew_{3} = 1 - (Q^\ast - Q) /Q^\ast
\end{equation}

\begin{equation}
    rew_{4} = 1 + (divergence^\ast - divergence) /divergence^\ast
\end{equation}

\begin{equation}
    score = rew_{T} = 10 \times (\alpha \times rew_{1} + \beta \times rew_{2}+ \gamma \times rew_{3} + \eta \times rew_{4})
\end{equation}

In the reward formulation, $rew_{1}$ involves several components. The target maximum to be achieved is denoted by $\lambda^\ast$, while $\lambda$ represents the current value of the wavelength obtained from the FDTD environment. To invert the reward and ensure that larger wavelength  result in larger rewards, a constant $1$ is used. Additionally, to normalize the magnitude of rewards, we use $\lambda^\ast$ as the denominator. In $rew_{2}$, the reward is defined based on  the target modal volume $area^\ast$. The formulation of $rew_{2}$ is defined such that rewards are higher when the calculated modal volume ($area$) are closer to $area^\ast$, which aligns with the objectives of the inverse design problem stated earlier. It is the same in $rew_{3}$ and $rew_{4}$, where we choose our target as emitting power (Q-factor) and divergence angle. Finally, Equation 8 defines the total reward $rew_{T}$ as a weighted sum from $rew_{1}$ to $rew_{4}$, which we also call score. Weighting coefficients are selected as follows after multiple rounds of tuning with different combinations: $\alpha = 1e+10, \beta = 1e+15, \gamma = 1e+30, \eta = 1e+32.$ We chose large coefficients because sometimes reward values could be as low as 1e-20 or even smaller. The target metrics are listed in Table~\ref{tab:pcsel_target_metric} below, and since we have an infinitely large simulation model with Bloch boundary conditions, we didn't report the lasing area, divergence and beam quality in this paper. This will be covered in future works.

\begin{table}[htbp]
\centering
\begin{tabular}{|c|c|}
\hline
\multicolumn{2}{|c|}{\textbf{Solved design parameters}}\\\hline
\textbf{State space} & \textbf{Solved Values} \\ \hline
substrate layer thickness & 0.1680 um  \\
n cladding layer thickness & 0.1460 um \\
active layer thickness & 0.1920 um\\
PhC layer thickness & 0 um \\
p cladding layer thickness & 0  um\\
refractive index of substrate layer & 0   \\
refractive index of n cladding /p cladding layer & 0.0150  \\
refractive index of active layer & -0.1000  \\
lattice spacing &  -0.0170 um  \\
hole radius factor& -0.1892   \\ \hline 
\end{tabular}
\caption{Solved design parameters of the optimized PCSEL.  Hole radius = hole radius factor $\times$ lattice spacing / 2}
\label{tab:pcsel_optimized_design_params}
\end{table}

\begin{table}[htbp]
\centering
\begin{tabular}{|c|c|}
\hline
\textbf{Metric/FOM}  & \textbf{Target values} \\ \hline
Operating wavelength & = 1310 nm  \\
Lasing area  & $\geq$ 0.36 $\mu m^2$  \\
Q factor & $\geq$ 10000 \\
Divergence angle  & $\leq$ $3^o$    \\  
Beam quality $M^2$ & $\leq$ 3  \\ \hline
\end{tabular}
\caption{Target Metric/FOM of the PCSEL device to be satisfied via optimization, including the Q-factor, lasing area, operating wavelength, beam quality, and divergence angle. An ideal PCSEL has the following characteristics: single-mode, high beam quality $M^2$, large emission area, and small divergence angle. Since $M^2$ is dependent on divergence angle, we didn't set it as a reward parameter.}
\label{tab:pcsel_target_metric}
\end{table}

\textbf{Optimization results} The best design parameters of the optimized PCSEL structure is summarized in Table~\ref{tab:pcsel_optimized_design_params} and the corresponding solved optical attributes are listed in Table~\ref{tab:llm4pcsel_results} in the main text. Please note that state variables are net changes in the design parameters, rather than the design parameters themselves. So a state variable equal to 0 means that there is zero change in that particular parameter. These optimized values can be used to fabricate a PCSEL device in clean-room with enhanced performance metrics.

\textbf{Calculations of output power-to-injecting power ratio}
To calculate the output power-to-injecting power ratio of our PCSEL, which is equivalent to the electron (photon)-to-photon conversion efficiency, we used the following formula: Poynting vector divided by dipole source power (the Poynting vector was calculated by Lumerical FDTD's near field power monitor, while the dipole power was set to 3.98265e-14 w in Lumerical FDTD), or,

\begin{equation}
    power\_ratio = Poynting / dipole\_power = Poynting / 3.98265\times 10^{-14}
\end{equation}

It should be noted that the power ratio calculated here is an ideal/theoretical value, and actual experimental results will be normally worse than this due to optical losses and heat dissipations.

\subsection{Appendix C: Choice of lasers and computation resources} 

\textbf{Choice of laser cavities for inverse design.}
Traditional VCSELs are lasers that emit light vertically from the surface of the semiconductor structure, allowing for efficient coupling with optical fibers and other optical components. Photonic crystals are artificial structures with periodic refractive index modulation in one, two, or three dimensions. This periodicity generates bandgaps, band edges, and other unique properties that determine the propagation characteristics of light at specific frequencies.

PCSELs are a type of vertical-cavity surface emitting laser, compared to traditional VCSELs, that utilizes two-dimensional photonic crystals to control multi-directional diffraction, resulting in single-mode, high-power, and low-divergence angle emission.

The basic design of a PCSEL includes a photonic crystal layer, an active layer, and several cladding layers, including p-n junctions and electrodes, as shown in Figure~\ref{fig:pcsel}. The photonic crystal layer typically serves as a resonance cavity. The active layer is usually composed of III-V materials (such as InP, GaAs, GaN, etc.) and is doped with materials to form quantum wells or quantum dot structures, enhancing emission efficiency and controlling emission characteristics. The cladding layers are doped with impurity atoms to form p-type or n-type semiconductor materials, increasing the carrier concentration and enhancing the material's electrical conductivity. 

When an electrical current is applied to the device, the carriers undergo carrier population inversion between the valence band and the conduction band in the active layer, leading to the phenomenon of population inversion. Subsequently, carrier recombination occurs, releasing photons. These photons further couple into the photonic crystal cavity, enhancing stimulated emission. The design of the photonic crystal layer determines the coupling strength, wavelength, and direction of the emitted light, making the proper design of the photonic crystal layer crucial for the overall quality of PCSELs.

\textbf{Computing resources and software packages used.} 
The RL code was meticulously developed in Python, strictly adhering to the algorithmic model depicted in the Figure~\ref{fig:L2DO}. 
Throughout the implementation, widely used machine learning libraries like PyTorch, OpenAI Gym, and Ray RLlib played a crucial role. In particular, Gym and RLlib proved to be particularly advantageous in accelerating progress. For the training of L2DO, we utilized two Dell workstations with 8 Intel Xeon Gold 5222 cores and an NVIDIA Quadro P4000 GPU. Additionally, comparative computations were performed on a cluster machine with 30 Intel Xeon Gold 5218 processors and 6 NVIDIA 2080Ti graphics cards. The cluster machine exhibited approximately 200\% higher computational efficiency compared to the Dell workstations.
 
Regarding the FDTD simulations, FDTD (Finite Difference Time Domain) is a numerical method that employs central difference quotients to replace the first-order partial derivatives of the field with respect to time and space. By recursively simulating the wave propagation process in the time domain, the FDTD method obtains the field distribution. This approach facilitates a more straightforward and efficient analysis of the wave propagation process.

\subsection{Appendix D: Experiments with Llama2} 

\begin{figure}[htbp]
  \centering
  \subfloat[]{\includegraphics[width=0.45\textwidth]{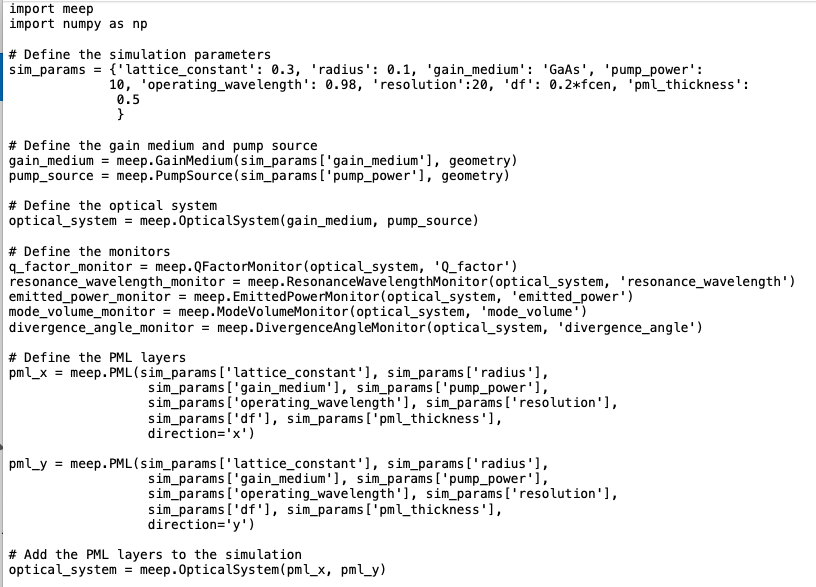}}
  \hfill
  \subfloat[]{\includegraphics[width=0.45\textwidth]{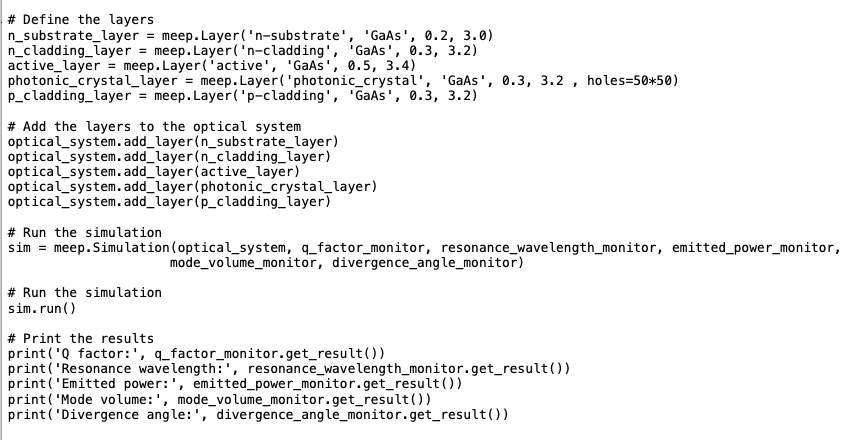}}
  \caption{Meep code generated by Llama2.}
  \label{fig:llama2_meep}
\end{figure}

\begin{figure}[htbp]
  \centering
  \includegraphics[width=0.95\linewidth]{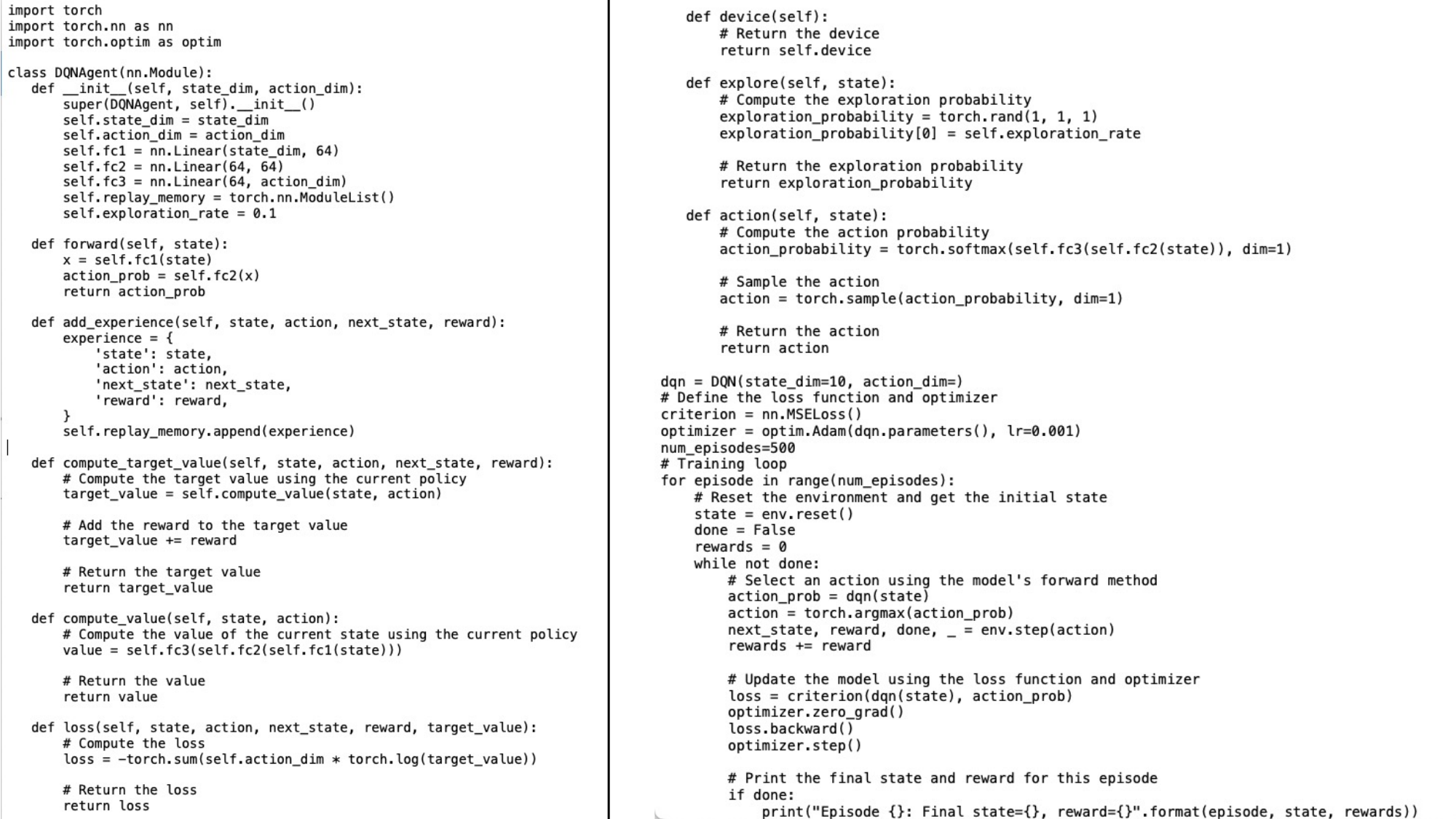}
  \caption{DQN code generated by Llama2.}
  \label{fig:llama2_dqn}
\end{figure}

Meep and DQN code generated by meta Llama2 is shown in Figure~\ref{fig:llama2_meep} and \ref{fig:llama2_dqn}, respectively. Compared to those generated by GPT in the main text, one can readily conclude that Llama2 is not nearly as powerful and capable as GPT-4. 
We can tell that both code are missing core components/modules that render the code erroneous or unable to execute. 
Critically, Llama2 doesn't seem to have the knowledge of meep and had huge trouble with writing correct methods/functions in meep. Sometimes it would even make up fake methods that doesn't exist at all. Moreover, Llama2 was unable to effectively correct the errors we fed back to it, and the human facilitator had to manually correct the errors. All in all, Llama2 proved to be inferior to GPT4 in virtually all aspects and for those who wish to efficiently design hardware with LLM, GPT is still the top choice. This, however, should not discourage anyone to use Llama2 because one can still harness the open-source power of Llama2.


The values of most of the hyperparameters were selected by performing an informal search with the PCSEL. 
We did not perform a systematic or exhaustive grid search owing to the high computational cost, although it is understandable that even better results could be obtained by systematically tuning the hyperparameter values one-by-one.

\subsection{Appendix E: Full comparison with literature data} 
We conducted an extensive literature survey of existing state-of-the-art (SOTA) PCSELs and summarized the comparison between our results and a selected group of papers in Table~\ref{tab:full_comparison} below (only square lattice and circular holes are reported).


\begin{table}[h]
    \centering
    \scriptsize
    \begin{tabular}{|c|c|c|c|c|c|c|c|c|c|c|c}
        \hline
         Paper & \makecell{lambda \\(nm)} & \makecell{PhC\\ dimension\\ (um)} & \makecell{Device\\area\\(um$^2$)}  & \makecell{Lattice\\ constant\\(nm)}  & Divergence & Q factor $\uparrow$ & \makecell{Lasing area\\ (m$^2$)} & M$^2$ & \makecell{half\\ theta} & \makecell{Loss\\ (1/nm)} $\downarrow$ \\
         \hline
         \citet{zhou2020continuous}  & 957 & 50*50 & 2500 & / & / & / & / & / & / & / \\
         \hline
         \citet{ohnishi2004room} & 959.44 & 50*50 & 2500 & / & $1.1^\circ$ &/ & 2.83e-09 &/ & 0.55 & / \\
         \hline
         \citet{sakai2005lasing} & 965 & 50*50 & 2500 & 286.25 & $1^\circ$ & 1700 & / & / & 0.5 & 1.29052e-05 \\
         \hline
         \citet{hsu2017electrically} & 1299 & 300*300 & 90000 & 390 & $\leq 2^\circ$ & 5000 & 1.76625e-08 & / & 1 & 3.22051e-06 \\
         \hline
         \citet{chen2017photonic} & 1260 & 300*300 & 90000 & / & / & / & / & / & / & / \\
         \hline
         \citet{chen2021improvement} & 948 & 125*125 & 15625 & 281 & $0.75^\circ$ & 2900 & 6.22e-09 & 3.1 & 0.375 & 7.70647e-06 \\
         \hline
         \citet{wang2021photonic} & 935 & 340*340 & 115600 & / & $0.38^\circ$ & / & / & 6.5 & 0.19 & / \\
         \hline
         \citet{reilly2020epitaxial} & 1010 & 250*250 & 62500 & / & / & / & / & / & / & / \\
         \hline
         \citet{kalapala2022scaling} & 1040 & 2000*2000 & 4e+6 & / & / & 100000 &  3.14e-08 & / & / & / \\
         \hline
         \textbf{Ours} & 1310 & 2*2 & 4 & 400 & $1.2^\circ$ & 36400 & 9.92e-14 & 1.36 & 0.6 & \textbf{4.30787e-07} \\
         \hline
    \end{tabular}
    \caption{Comparative assessment of SOTA PCSELs, where GaAs serves as the gain material and PhC lattices are characterized by square and circular holes. The "/" indicates the absence of data in the cited literature.}
    \label{tab:full_comparison}
\end{table}